# The Law of Closest Approach

M.N. Tarabishy

goodsamt@gmail.com

**Abstract:**

In this work, we introduce the Law of Closest Approach which is derived from the properties of conic orbits and can be considered an addendum to Kepler's laws. It states that: On the closest approach, the distance between the objects is minimal and the velocity vector is perpendicular to the position vector with maximum speed. The ratio of twice the kinetic energy to the negative potential energy is equal to the eccentricity plus one, or: $-2K/U = 1+e$. The advantage of this law is that both quantities (speed and position) are at extremum making the calculation of the eccentricity more robust.

## 1. Introduction:

The book of Ptolemy Almagest written around 150 AD was the standard text and the reference in astronomy during the Middle Ages. The book was translated into Arabic in the 8th century, and back to Latin in the 12th century. Its basic ideas are that the earth is a fixed sphere around which everything revolve in perfect circles, however, to account for the actual motion additional circular motions were added on top of the circular motion (using epicycles and deferents) making the model complicated but gave reasonably good predictions.

In 1532 shortly before his death, Copernicus, an astronomer, a mathematician, and a Catholic Canon, published his book "De Revolutionibus Orbium Coelestium". The book argues for a Heliocentric view of the world instead of the Geocentric model of Ptolemy that was the standard model for the heavens for over a thousand years and a religious dogma that the church vigorously defended.

Copernicus' ideas and his book caused a lot of religious and scientific controversy and sparked the scientific revolution. Removing the earth from the center of the universe gave rise to the basic principle in astronomy that there is no preferred place in the universe. While Heliocentric model was simpler and did not have the problem of retrograde motion that the Geocentric model had to deal with, it was not completely correct as Copernicus thought the orbits were circles and therefore, his model did not provide better predictions than Ptolemy's much more complex model, and therefore, more work was needed.

With good tools and careful observations, Tycho Brahe, the Danish astronomer, accumulated the best data of his time, and upon his sudden death in 1601, his German employee, Kepler, an astronomer and a mathematician, inherited his data.

Kepler has found his laws empirically and after great efforts in trying to fit the available data to the best orbit shape and relations. He published his first two laws in "Astronomia Nova" in 1609, while the third law appeared in his book "Harmonices Mundi" in 1619. His three laws are foundational in astronomy to understand the motion of objects in the heavens [6].

**I**- All planetary orbits are ellipses with the sun in one focus.

**II**- Planets sweep out equal areas in equal times.

**III**- The squares of the orbital periods are proportional to the cubes of the distances from the sun.

While Kepler's efforts were successful, it was Newton in his monumental book: Mathematical Principles of Natural Philosophy "Principia" published in 1687 who established the theoretical basis for Kepler's laws and derived them from his laws of gravitation and motion [5].

In the next section we present the derivation of the laws of motion of two bodies that interact through the gravitational forces between them.

## 2. The Two Body Problem:

In the following we derive the equations of motion using the method used in the book of Bates et al, [1].

We have two masses: m, and M that are separated by a distance r. We establish an inertial reference frame, XYZ, and another non inertial frame xyz with its center at the center of mass M as in figure 1.

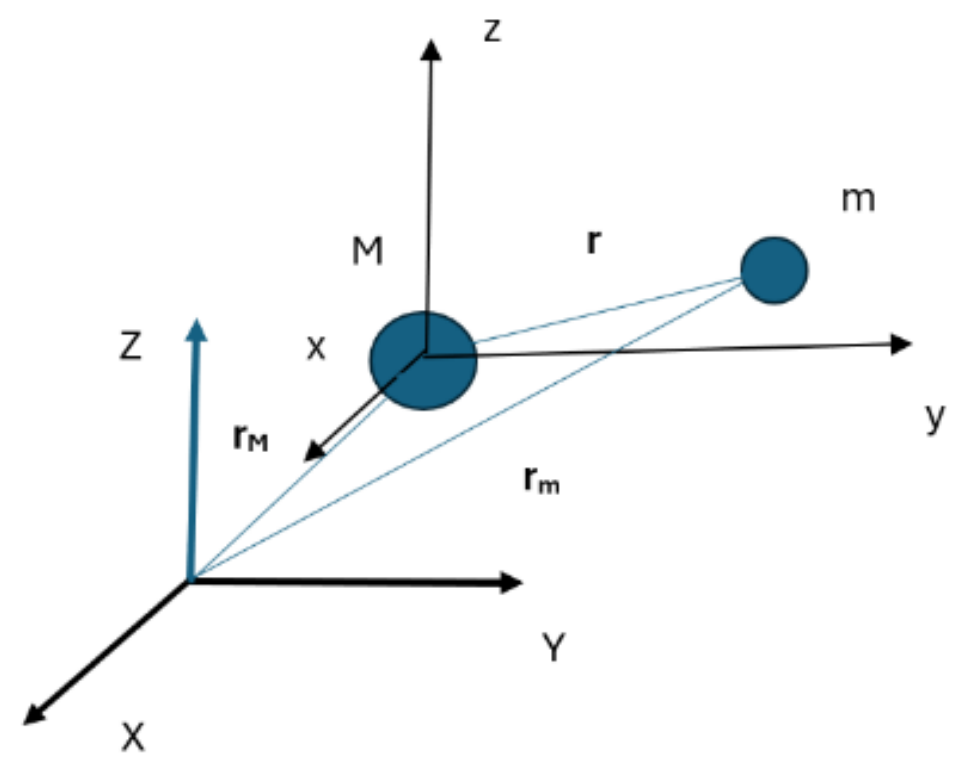

Figure 1: Two masses position representation

Using Newton second law of motion, the force affecting the mass m as a result of mass M attraction is equal to the acceleration times the mass m is equal to gravitational force and given in equation (1):

$$\ddot{\boldsymbol{r}}_{\boldsymbol{m}} = \frac{-\,G\,M}{r^3}\,\boldsymbol{r} \qquad (\mathbf{1})$$

Also, the equation of motion for mass M is given by:

$$\ddot{\boldsymbol{r}}_{\boldsymbol{M}} = \frac{G\,m}{r^3}\,\boldsymbol{r} \qquad (\mathbf{2})$$

Subtracting equation (2) from (1), and arranging:

$$\ddot{\boldsymbol{r}}_{\boldsymbol{m}} - \ddot{\boldsymbol{r}}_{\boldsymbol{M}} = \ddot{\boldsymbol{r}} = \frac{-\,G\,(m+M)}{r^3}\,\boldsymbol{r} \qquad (\mathbf{3})$$

Or,

$$\ddot{\boldsymbol{r}} = \frac{-\,G\,(m+M)\boldsymbol{r}}{r^3} = \frac{-\mu}{r^3}\,\boldsymbol{r} \qquad (\mathbf{4})$$

Giving:

$$\ddot{\boldsymbol{r}} + \frac{\mu}{r^3}\,\boldsymbol{r} = \mathbf{0} \qquad (\mathbf{5})$$

Where:

r: is position of mass m w.r.t. mass M

μ: The gravitational parameter μ = G (m+M), or μ ≈ G M if m << M.

Equation (5) is the vector equation of motion for two objects interacting through the force of gravity. Dot product both sides by the velocity vector:

$$\dot{\boldsymbol{r}}\,.\,\left(\ddot{\boldsymbol{r}} + \frac{\mu}{r^3}\,\boldsymbol{r} = \mathbf{0}\right) \qquad (\mathbf{6})$$

We get the equation:

$$\frac{d}{dt}\left(\frac{V^2}{2}\right) + \frac{d}{dt}\left(\frac{-\mu}{r}\right) = 0 \qquad (\mathbf{7})$$

We define the specific total mechanical energy ε:

$$\varepsilon = \frac{V^2}{2} - \left(\frac{\mu}{r} + c\right) \qquad (\mathbf{8})$$

Setting the constant of integration c to zero: c=0, so that the specific potential energy: U = -μ/r = 0 @ r = ∞, and ε = constant, and the total mechanical energy ε is conserved.

$$\varepsilon = \frac{V^2}{2} - \frac{\mu}{r} \quad \mathbf{(9)}$$

Cross multiplying equation (5) by relative position vector **r**:

$$\boldsymbol{r} \times \left( \ddot{\boldsymbol{r}} + \frac{\mu}{r^3}\,\boldsymbol{r} = \boldsymbol{0} \right) \quad \mathbf{(10)}$$

Manipulating and rearranging gives:

$$\frac{d}{dt}\,(\boldsymbol{r} \times \dot{\boldsymbol{r}}) = \frac{d}{dt}(\boldsymbol{r} \times V) = \frac{d}{dt}\,(\,\boldsymbol{h}\,) = \boldsymbol{0} \quad \mathbf{(11)}$$

With **h**: The specific angular momentum vector.

**h** is constant indicates conservation of angular momentum, and the motion is in a fixed plane.

$$\boldsymbol{h} = (\boldsymbol{r} \times \boldsymbol{V}) = \boldsymbol{const.} \quad \mathbf{(12)}$$

With the magnitude:

$$h = r\,V\,sin(\gamma) = r\,V\,cos(\phi) \quad \mathbf{(13)}$$

ɣ: Angle between position vector **r**, and velocity vector **V**.

ф: Flight path angle (angle between velocity vector and local horizon).

The vector **h** is perpendicular to the plane of **(r,V).**

Crossing the equation of motion (5) by the vector **h**:

$$\left( \ddot{\boldsymbol{r}} = -\frac{\mu}{r^3}\,\boldsymbol{r} \right) \times \boldsymbol{h} \quad \mathbf{(14)}$$

Gives:

$$\frac{d}{dt}\,(\dot{\boldsymbol{r}} \times \boldsymbol{h}\,) = \frac{-\mu}{r^3}(\,\boldsymbol{r} \times \boldsymbol{r} \times \boldsymbol{V}) = \frac{-\mu}{r^3}\,[\,\boldsymbol{r}\,(\,\boldsymbol{r}\,.\boldsymbol{V}\,) - \boldsymbol{V}\,(\,\boldsymbol{r}\,.\boldsymbol{r}\,)] \quad \mathbf{(15)}$$

Or,

$$\frac{d}{dt}\,(\dot{\boldsymbol{r}} \times \boldsymbol{h}\,) = \frac{\mu}{r}\boldsymbol{V} - \frac{\mu\,V}{r^2}\,\boldsymbol{r} = \mu\,\frac{d}{dt}\left(\frac{1}{r}\,\boldsymbol{r}\right) \quad \mathbf{(16)}$$

Integrating:

$$(\dot{\boldsymbol{r}} \times \boldsymbol{h}\,) = \mu\,\left(\frac{1}{r}\,\boldsymbol{r}\right) + \boldsymbol{B} \quad \mathbf{(17)}$$

Where the integration constant **B** is a vector that points to the periapsis.

Dot multiplying the equation (17) by position vector **r**:

$$\boldsymbol{r} \,.\, \left[ (\dot{\boldsymbol{r}} \times \boldsymbol{h}) = \mu \left( \frac{1}{r} \boldsymbol{r} \right) + \boldsymbol{B} \right] \tag{18}$$

Using the property of scalar triple product and rearranging

$$r = \frac{\frac{h^2}{\mu}}{1 + \frac{B}{\mu} \cos(\upsilon)} = \frac{p}{1 + e \cos(\upsilon)} \tag{19}$$

With: $\upsilon$ is the angle between the vector **B** and position vector **r**.

$$p = \frac{h^2}{\mu} \tag{20}$$

Where: p is semi latus rectum.

The eccentricity vector:

$$\boldsymbol{e} = \boldsymbol{B} \frac{1}{\mu} \tag{21}$$

And

$$\boldsymbol{e} = \boldsymbol{B} \frac{1}{\mu} = \frac{1}{\mu} (\dot{\boldsymbol{r}} \times \boldsymbol{h}) - \frac{1}{r} \boldsymbol{r} \tag{22}$$

Giving:

$$\boldsymbol{e} = \frac{1}{\mu} \left[ \boldsymbol{r} \left( V^2 - \frac{1}{r} \right) - \boldsymbol{V} \, (\boldsymbol{V}.\boldsymbol{r}) \right] \tag{23}$$

Equation (23) gives the relation for obtaining the eccentricity vector for any conic orbit.

For any conic orbit (except the parabola because a = c = ∞)

$$e = \frac{c}{a} \tag{24}$$

$$p = a \, (1 - e^2) \tag{25}$$

Where a is the major semi radius.

$$e = \sqrt{1 - \frac{p}{a}} \tag{26}$$

For any conic orbit:

$$r_{min} = r_{periapsis} = \frac{p}{1 + e} \tag{27}$$

$$\varepsilon = -\frac{\mu}{2a} \qquad \textbf{(28)}$$

$$e = \sqrt{1 + \frac{2\varepsilon h^2}{\mu^2}} \qquad \textbf{(29)}$$

## 3. The Law of Closest Approach:

We follow similar steps as above, starting from the equation of motion (5), and we cross product both sides by vector **h**:

$$\left( \ddot{\boldsymbol{r}} = -\frac{\mu}{r^3}\boldsymbol{r} \right) \times \boldsymbol{h} \qquad \textbf{(30)}$$

Giving:

$$\frac{d}{dt}(\dot{\boldsymbol{r}} \times \boldsymbol{h}) = \frac{\mu}{r}\boldsymbol{V} - \frac{\mu V}{r^2}\boldsymbol{r} = \mu\frac{d}{dt}\left(\frac{1}{r}\boldsymbol{r}\right) \qquad \textbf{(31)}$$

Integrating, with the vector **B** as the constant of integration:

$$(\dot{\boldsymbol{r}} \times \boldsymbol{h}) = \mu\left(\frac{1}{r}\boldsymbol{r}\right) + \boldsymbol{B} \qquad \textbf{(32)}$$

Rearranging:

$$\frac{(\boldsymbol{V} \times \boldsymbol{r} \times \boldsymbol{V})}{\mu} = \hat{\boldsymbol{r}} + \boldsymbol{B}\frac{\mathbf{1}}{\mu} \qquad \textbf{(33)}$$

Then, we dot product both sides by unit position vector $\hat{\boldsymbol{r}}$ :

$$\hat{\boldsymbol{r}} \,.\, \left\{\frac{1}{\mu}\left[\boldsymbol{r}\,(\boldsymbol{V}.\boldsymbol{V}) - \boldsymbol{V}\,(\boldsymbol{V}.\boldsymbol{r})\right] = \hat{\boldsymbol{r}} + \boldsymbol{B}\frac{1}{\mu}\right\} \qquad \textbf{(34)}$$

If we choose the case where **V** ⊥ **r** (For any conic orbit, at the periapsis or apoapsis, the velocity vector **V** is perpendicular to the position vector **r**, and both vectors have an extremum magnitude), then, we get:

$$\boldsymbol{\frac{2K}{-U} = \frac{V^2}{\frac{\mu}{r}} = 1 + e} \qquad \textbf{(35)}$$

For a circle: e = 0.  For an ellipse e < 1.  For a parabola e = 1.  For a hyperbola e >1.

In figure 2 we show the vector **e** and the closest point or periapsis.

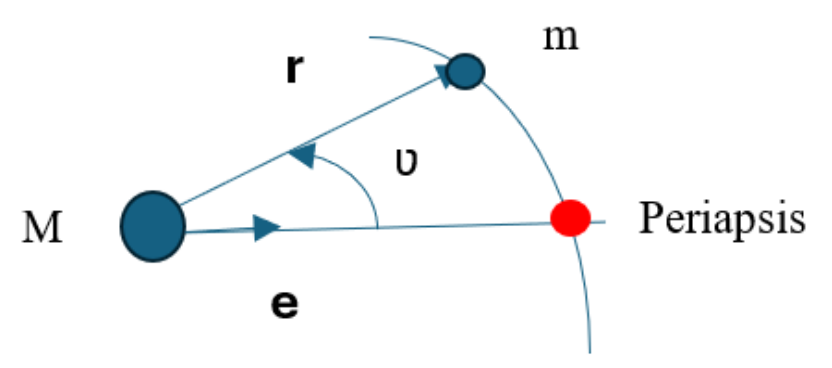

Figure 2. Vector **e** and the periapsis

We have searched the literature but could not find a relation similar to equation (35).

### 3.1. Examples:

In the following, we pick several planets and comets, with their closest approach velocity, distance as well as their masses (v_p, r_p, M). Our calculation (Orbit_e) of eccentricity e using eqn. (35), and the last column is the data publicly available from NASA [8].

| Object | v_p [m/s] | r_p [m] | M [kg] | -2K/U | Orbit_e | Pub_Data |
|---|---|---|---|---|---|---|
| Earth | 3.03E+04 | 1.47E+11 | 5.9722E+24 | 1.02E+00 | 0.01692 | 0.0167 |
| Mars | 2.65E+04 | 2.06E+11 | 6.42E+23 | 1.09E+00 | 0.090058 | 0.0934 |
| Mercury | 5.90E+04 | 4.60E+10 | 3.30E+23 | 1.21E+00 | 0.205343 | 0.2056 |
| Venus | 3.53E+04 | 1.07E+11 | 4.8673E+24 | 1.01E+00 | 0.006892 | 0.0068 |
| Pluto | 6.10E+03 | 4.43E+12 | 3.01E+22 | 1.24E+00 | 0.243491 | 0.2444 |
| Halley | 9.95E+02 | 5.26E+12 | 2.2E+14 | 3.92E-02 | -0.96076 | 0.967 |
| Omuamua | 8.77E+04 | 3.88E+10 | 4.00E+06 | 2.25E+00 | 1.247027 | 1.2011 |
| 2I/Borisov | 4.92E+04 | 2.99E+11 | 2.00E+14 | 5.45E+00 | 4.45 | 3.36 |

For all cases we used the periapsis (closest point), however, for Halley's comet we used the apoapsis (the furthest point) and we got the value of e negative. We kept the negative sign in the table to draw attention to the issue. The calculated values and the ones obtained from the public domain are very close except for 2I/Borisov comet. The difference might be due to the gas ejection as it approaches the sun.

### 3.2 Benefits:

The distance is reaching a minimum, therefore, it is less sensitive to measurement errors.

Also, the velocity is at maximum and therefore, it is less sensitive to measurement errors.

Calculating the eccentricity is very simple, and therefore, identification of the path is simplified.

We can also give some insight into the degenerative or rectilinear conic sections (ellipse, parabola, and hyperbola).

If velocity vector **V** is along the apse line towards the periapsis, then, **h** = **0**, and eqn. (32) gives:

$$\mathbf{0} = \mu\left(\frac{1}{r}\boldsymbol{r}\right) + \boldsymbol{B} \qquad (36)$$

Then, if divide by μ and we dot the equation with unit vector $\hat{\boldsymbol{r}}$ , we get always:

$$|e| = 1 \qquad (37)$$

If we shoot an object upwards, the specific energy needed to take an object from say the surface of the earth (with mass M and radius R) to infinity is equal to the work against the force of gravity, and is given by:

$$E_{es} = U_{es} = \frac{G\,M}{R} \qquad (38)$$

This energy needs to be provided as kinetic energy $K_o = 0.5\ V^2$ at the launch of the object, and therefore, equation (35) tells us that we need $K_o = U_{es}$ , and it corresponds to the case of a parabola that degenerated to a straight line.

If $K_o < U_{es}$ , then we have a line that corresponds to a degenerate ellipse and the object will fall back to earth, while if $K_o > U_{es}$ is a line that corresponds to a degenerative hyperbola.

### 3.3 Relation to Virial Theorem:

The virial theorem relates for a system of objects (points) with central forces F of the form:

$$F \propto r^n \qquad (39)$$

Where r is the position vector with the average kinetic energy <K> and average potential energy <U> of the system by the relation [7]:

$$< K > = 0.5\,(n+1) < U > \qquad (40)$$

With the square inverse law taking the form:

$$< K > = -0.5\ < U > \qquad (41)$$

And we notice that it is a very similar relation that we got in equation (35).

For example, the virial theorem can be used to estimate the mass of a cluster of N stars (or galaxies) assuming a spherical symmetry, with each having a mass of m, then the total mass will be:

$$M = N.m$$

And the estimated average potential energy:

$$< U > \approx G\,M^2/R$$

Where R is the radius of the cluster. The kinetic energy is given by:

$$< K > N.\frac{m < V >^2}{2} = \frac{M < V >^2}{R}$$

Therefore, the estimated mass is:

$$M \approx \frac{R < V >^2}{G}$$

## 4. Conclusion:

In this work, we derived a relation for the closest approach of a conic orbit, and we called it: "The Law of Closest Approach" and it states that: On the closest approach (at periapsis), the distance between the objects is minimal and the velocity vector is perpendicular to the position vector with maximum speed. The negative ratio of twice the kinetic energy to the potential energy is equal to the eccentricity plus one, or: -2K/U = 1 + e.

For the case of Vp (the speed at periapsis radial), then, e = 1, and K = U, and the kinetic energy needed is equal to the potential energy, and we the object leaves the planet (degenerate parabola).

The benefit of this relation is that it is simple to use to find the eccentricity, in addition to the fact that both the velocity and the position of the object have extremum and that makes the quantities less sensitive to errors.

We have also shown a relationship of our relation to the Virial Theorem, where the relation has the same form as that given by the virial theorem.